\documentclass[%
reprint,
showpacs,preprintnumbers,
 amsmath,amssymb,
 aip,
 graphicx,
apl,
]{revtex4-1}

\usepackage{float,graphicx}% Include figure files
\usepackage{dcolumn}% Align table columns on decimal point
\usepackage{bm}% bold math
\usepackage[mathlines]{lineno}% Enable numbering of text and display math
\usepackage{ulem}
\usepackage{color}
\usepackage[version=3]{mhchem}

\begin{document}

\preprint{Accepted in AIP Advances}

\title[Short Title]{The effect of ion irradiation on dephasing of coherent optical phonons in GaP} 

\author{Takuto Ichikawa}
\affiliation{Department of Applied Physics, Graduate school of Pure and Applied Sciences, University of Tsukuba, 1-1-1 Tennodai, Tsukuba 305-8573, Japan}
\author{Yuta Saito}
\affiliation{Device Technology Research Institute, National Institute of Advanced Industrial Science and Technology, Tsukuba Central 5, 1-1-1 Higashi, Tsukuba 305-8565, Japan.}
\author{Muneaki Hase}
\email{mhase@bk.tsukuba.ac.jp}
\affiliation{Department of Applied Physics, Faculty of Pure and Applied Sciences, University of Tsukuba, 1-1-1 Tennodai, Tsukuba 305-8573, Japan}

\date{\today}

\begin{abstract} 
The dephasing of coherent longitudinal optical (LO) phonons in ion-irradiated GaP has been investigated with a femtosecond pump-probe technique based on electro-optic sampling. 
The dephasing time of the coherent LO phonon is found to be dramatically prolonged by the introduction of a small amount of defects by means of 
Ga-ion irradiation. The maximum dephasing time observed at room temperature is 9.1 ps at a Ga$^{+}$ ion dose of 10$^{13}$/cm$^{2}$, which is significantly longer 
than the value of 8.3 ps for GaP before ion irradiation. The longer dephasing time is explained in terms of the suppression of electron-LO-phonon scattering by the presence of defect-induced deep levels. 
\end{abstract} 


\maketitle 

In the last two decades, the progress of femtosecond (fs) lasers has made possible a time-domain spectroscopy of the coherent lattice vibration in 
solids.\cite{Zeiger92,Shah99,Dekorsy00} 
The coherent phonons are impulsively generated by ultrashort light pulses, and lose coherence due to the interaction with their environment, e.g., 
other lower energy phonon modes.\cite{Hase10} There was active discussion regarding the generation mechanisms of the coherent phonons in 
semimetals,\cite{Zeiger92} transparent materials,\cite{Yan85} semiconductors,\cite{Dekorsy00,Yee02} and semiconductor multiple quantum 
wells.\cite{Yee01} Recently, there has been actively reported that driving coherent phonons induces ultrafast structural distortion in ferromagnet,\cite{Forst11} antiferromagnet,\cite{Juraschek17} and phase-change materials.\cite{Hase15} 
However, little is known how coherent phonons lose their coherence (dephase) in defective semimetals and semiconductors.  

GaP is a III-V semiconductor with an indirect band-gap of $\approx$2.25 eV at room temperature (RT) and promising material for device applications such as light emitting diodes.\cite{Henry76,Wight77} 
In addition, GaP has the largest refractive index ($n$ $>$ 3), enabling strong optical confinement and implying a large $\chi^{(3)}$ nonlinearity. 
The non-centrosymmetric crystal structure of GaP yields a nonzero piezo-electric effect and large $\chi^{(2)}$ nonlinearity.\cite{Wilson20}
%The dynamics of photo-generated carriers has been extensively investigated with fs pulse lasers.\cite{Shah99} 
The coherent longitudinal optical (LO) phonon in GaP was observed by a transient reflectivity technique.\cite{Chang03} 
It has been proposed that under above band-gap excitation in GaAs the coherent LO phonons are generated by a sudden screening of the surface space-charge field (SCF) 
by the photo-excited carriers.\cite{Dekorsy00} 
On the other hand, under the condition of below band-gap excitation, this is the case here, impulsive stimulated Raman scattering (ISRS) will dominate for the generation of coherent LO phonons,\cite{Yan85,Yee02,Ishioka15} while phonon dephasing by multiphoton-generated carriers is possible.\cite{Yee02}
At a sufficiently low photo-excited carrier density, the dephasing process of the coherent LO phonon in GaP has been accounted for 
anharmonic decay due to phonon-phonon interaction,\cite{Bron86} and the dephasing time was obtained to be $\sim$ 10 ps at room 
temperature.\cite{Bron86} 
By increasing the photo-excited carrier density, the dephasing time was decreased to below 1 ps. \cite{Ishioka15} This suggests that 
another dephasing process related to electron-phonon interaction is operative at a high carrier density.\cite{Smith92}

Incorporation of a small amount of defects into semiconductors is an intriguing method for manipulating their electrical properties 
such as carrier lifetime\cite{Gupta92} and carrier mobility.\cite{Yu99} 
Ultrafast carrier dynamics have been examined in low-temperature (LT) grown GaAs, which has been a key material for the generation of THz 
radiation since 1990's.\cite{Tani97} 
The carrier lifetime was  decreased to a sub-picosecond time scale due to trapping of carriers by As-precipitates.\cite{Gupta92,Benjamin96} Defects 
also affect the lattice properties in solids, e.g., the dephasing time and the frequency of phonons.\cite{Dekorsy93} 
In fact, the dephasing time of the coherent phonons was found to decrease with increasing density of lattice defects, which was explained 
by phonon-defect scattering.\cite{Hase00} Introduction of lattice defects can thus be a perturbation of elementary excitations in solids through 
localized defect potential.\cite{Hase10} 
Furthermore, introduction of lattice defects can change the nonlinear optical effect through nonlinear susceptibility, e.g., $\chi^{(2)}$ and $\chi^{(3)}$.\cite{Motojima19}

In this paper, the effect of lattice defects on the dephasing of the coherent LO phonon in GaP is investigated. The dephasing time of the 
coherent LO phonons is found to be dramatically prolonged by the introduction of a small amount of vacancies by means of ion irradiation. 
The prolongation is not expected by the phonon-defect scattering, but is explained in terms of a modification of decay channel of photo-generated 
carriers by lattice defects, resulting in the suppression of electron-phonon scattering. 

\begin{figure}
\includegraphics[width=7.0cm]{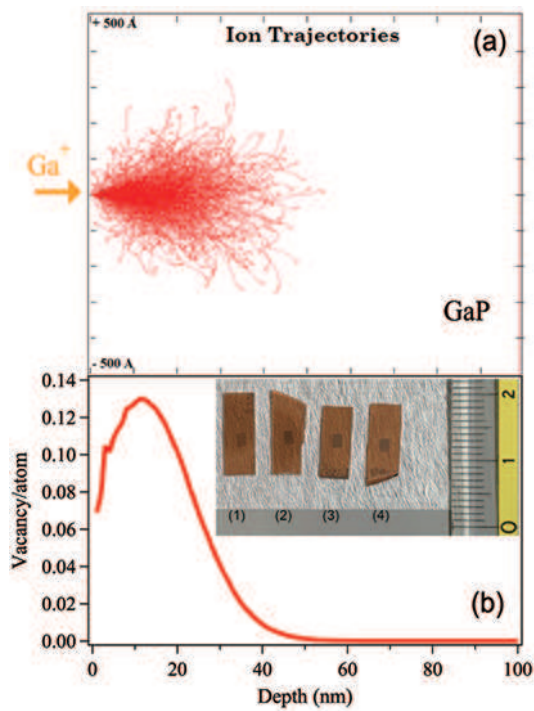}
\caption{(a)  Monte Carlo simulations of Ga ion irradiation.(b) The profile of the vacancy introduced by Ga$^{+}$ ion irradiation.  Inset: Photograph of the Ga ion irradiated GaP samples using FIB for (1) 1.0$\times$10$^{13}$Ga$^{+}$/cm$^{2}$, (2) 2.0$\times$10$^{13}$Ga$^{+}$/cm$^{2}$, (3) 4.0$\times$10$^{13}$Ga$^{+}$/cm$^{2}$, and (4) 8.0$\times$10$^{13}$Ga$^{+}$/cm$^{2}$.
}
\label{Fig1}
\end{figure}

The samples used were non-doped GaP (100) wafers with 300 $\mu$m thick. 
In order to introduce lattice defects in a controlled manner, the samples were irradiated with 30 keV Ga$^{+}$ ions at various doses from 1.0$\times$10$^{13}$ to 8.0$\times$10$^{13}$ Ga$^{+}$/cm$^{2}$ in a focused ion beam (FIB) system [Fig. \ref{Fig1}(a)]. 
The irradiation of the Ga$^{+}$ beam induces Ga- and P-vacancies.\cite{Legodi00} The damage profile was calculated by Monte Carlo 
simulations\cite{Ziegler85} to be a Gaussian function with a peak at $\approx$12 nm from the surface and a width of $\approx$16 nm, as shown in Fig. \ref{Fig1}(b).  

\begin{figure}
\includegraphics[width=8.0cm]{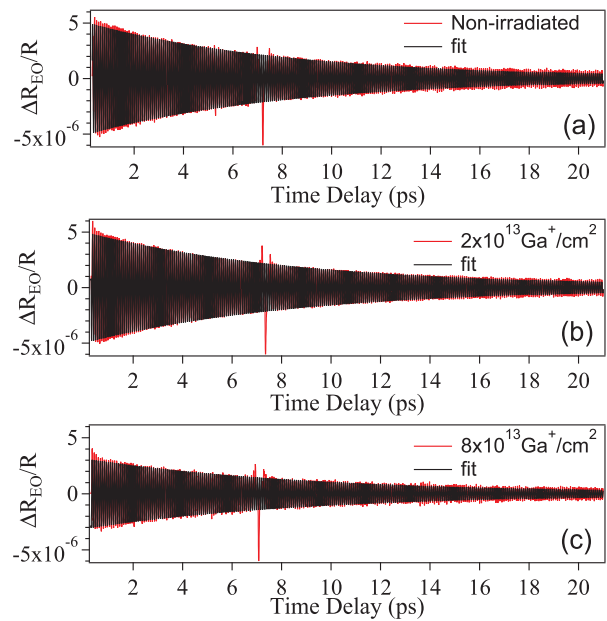}
\caption{The time domain signal of coherent LO phonons in GaP observed for (a) non-doped, (b) 2.0$\times$10$^{13}$Ga$^{+}$/cm$^{2}$, and (c) 8.0$\times$10$^{13}$Ga$^{+}$/cm$^{2}$. 
The black lines represent the fit using a damped harmonic oscillator. A transient signal appears at $\approx$7 ps is due to a back-reflection of the probe beam and does not affect the dephasing of the coherent LO phonon. 
}
\label{Fig2}
\end{figure}

Femtosecond pump-probe measurements were performed at 300 K. The light source used was 
a mode-locked Ti:sapphire laser with a pulse duration of $\approx$25 fs, a central wavelength of 830 nm ($\approx$1.49 eV), and a repetition rate of 80 MHz. 
The average powers of pump- and probe-beams were fixed at 175 mW and 8.7 mW, respectively. 
Both the pump- and probe-beams were focused via off-axis mirror to a diameter of $\sim$ 20 $\mu$m on the sample and the pump- and probe-fluences were kept at $\approx$ 1.19 m$ \mathrm{J/cm^2}$ and 60 $\mathrm{\mu J/cm^2}$, respectively. 
The optical penetration depth at 830 nm was estimated from the absorption coefficient to be $\approx$130 nm.\cite{Ishioka15}
Since the optical penetration is an exponential function, the optical penetration depth overlaps well with that of the vacancy distribution (Fig. \ref{Fig1}). 
Because of much less transition probability across the indirect band-gap at the X-valley, 
the one-pump and one-probe pulses can excite carriers via two-photon absorption (TPA) near the zone center ($\Gamma$ point), 
where the direct gap is $\approx$2.78 eV at RT.\cite{Ishioka15} 
Anisotropic reflectivity change ($\Delta R_{EO}/R$) was measured by an electro-optic (EO) sampling technique \cite{Dekorsy00} to detect coherent 
LO phonons as a function of the delay time. The delay between the pump and probe pulses was scanned by an oscillating retroreflector operated at a frequency of 10.5 Hz up to 30 ps (Ref. \cite{Hase12}). 

\begin{figure}
\includegraphics[width=7.0cm]{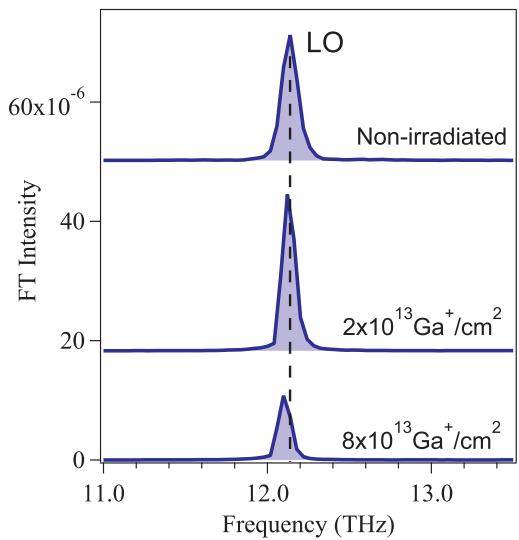}
\caption{ The FT spectra of the coherent LO phonon obtained for the different Ga$^{+}$ dose levels.  
The dashed line represents peak position of the coherent LO phonon before the Ga$^{+}$ irradiation. 
}
\label{Fig3}
\end{figure}

The time derivatives of the $\Delta R_{EO}/R$ signals for non-doped GaP before and after Ga$^{+}$ irradiation are shown in Fig. \ref{Fig2}. 
The coherent oscillation due to the LO phonon (12.15 THz) is observed in three different samples, as shown in the 
Fourier transformed (FT) spectra in Fig. \ref{Fig3}.\cite{Chang03,Ishioka15}
The peak frequency of the LO mode implies a significant change from $\approx$12.2 THz down to $\approx$12.1 THz as seen in Fig. \ref{Fig3}, which will be discussed in more details in the later section. Note that in Fig. \ref{Fig3} the peak intensity of the LO mode for 2.0$\times$10$^{13}$Ga$^{+}$/cm$^{2}$ is slightly greater and the width is narrower than that of the non-irradiated GaP, implying the dephasing time of the LO mode becomes longer for 2.0$\times$10$^{13}$Ga$^{+}$/cm$^{2}$ than that before the Ga$^{+}$ irradiation. 

\begin{figure}
\includegraphics[width=7.5cm]{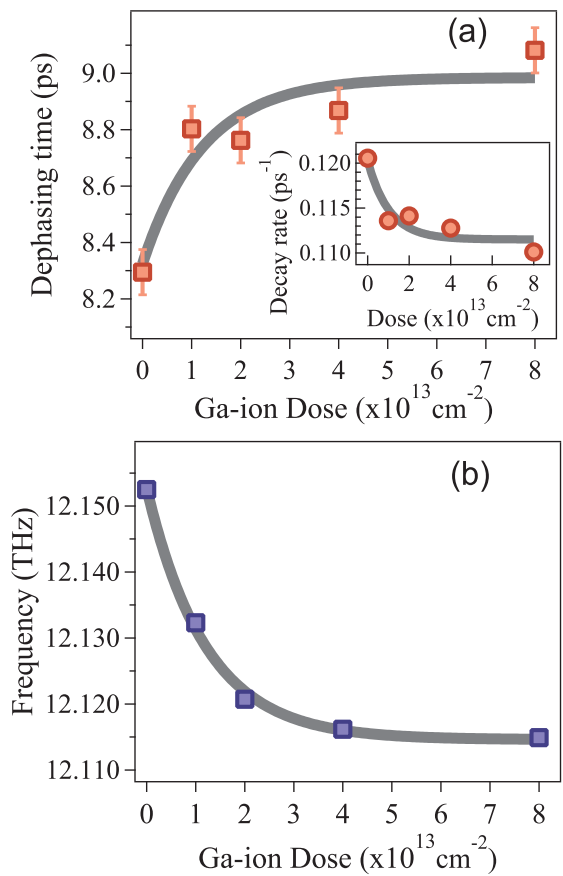}
\caption{(a) The dephasing time of the coherent LO phonon as a function of Ga$^{+}$ dose. 
The inset represents the decay rate, the inverse of the dephasing time. (b) The frequency of the coherent LO phonon as a function of the density of Ga$^{+}$ dose. 
The solid curve represents guide for the eyes. 
}
\label{Fig4}
\end{figure}

To further investigate the effect of the irradiation on the dephasing of the coherent LO phonon, the time-domain data were analyzed with an exponentially damped harmonic oscillation as shown in Fig. \ref{Fig2}. 
The dephasing time of the coherent LO phonon thus obtained is plotted in Fig. \ref{Fig4}(a) as a function of Ga$^{+}$ dose. The frequency of the LO phonon is also plotted in Fig. \ref{Fig4}(b) as a function of Ga$^{+}$ dose. 
The frequency gradually shifts from $\approx$12.15 down to $\approx$12.11 THz at the irradiation of 8.0$\times$10$^{13}$ Ga$^{+}$/cm$^{2}$.
The decrease of the peak frequency of $\sim$0.04 THz is possibly governed by strains introduced by the implanted Ga$^{+}$ ions, i.e., the strains resulting from the displacement of the host Ga and P lattice from its normal equilibrium position.\cite{Rao1983}
The dephasing time before ion-irradiation was 8.3 $\pm$ 0.1 ps, which is comparable but significantly shorter than that obtained using 800 nm light 
for $n$-doped GaP Schottky diode, $\sim$10 ps (Ref. \cite{Chang03}), and $n$-doped GaP wafer, $\sim$12 ps (Ref. \cite{Ishioka15}). 
The shorter dephasing time before the irradiation implies the existence of electron-phonon interaction under the presence of photo-generated carriers, as discussed later in the present study.\cite{Ishioka15}
Irradiation of 8.0$\times$10$^{13}$ Ga$^{+}$/cm$^{2}$ $increases$ the dephasing time up to 9.1 ps, being consistent with the peak narrowing observed in the FT spectra (see after irradiation in Fig. \ref{Fig3}). 
 
Scattering of the coherent phonon by lattice defects generally promotes its dephasing when the ion dose was typically 
$\geq$ 10$^{14}$/cm$^{2}$ (Ref. \cite{Hase00}). 
However, the $increase$ in the dephasing time for lower ion doses cannot be explained by defect scattering. 
Dekorsy and coworkers observed a slightly longer dephasing time for the coherent LO phonon in as-grown 
LT-GaAs (4.0 ps) than in annealed LT-GaAs (3.5 ps), in which better crystal quality serves longer dephasing time. 
They explained the longer dephasing time in terms of phonon localization,\cite{Dekorsy93} but the mechanism would 
lead only to a decrease in the dephasing time.\cite{Hase00} 
Even if there were structural modifications, it would lead only to a decrease in the dephasing time, corresponding to the broadening of Raman 
line in defective GaAs,\cite{Kitajima97} being contrary to the present results. 

The prolonged dephasing time can be explained by considering the (one)electron-LO phonon coupling,\cite{Hase03,Dharmarasu} or plasmon-LO phonon coupling.\cite{Vallee97,Kunugita15}
The decay rate 1/$\tau$ of the coherent LO phonon in defective GaP is given by the sum of the intrinsic anharmonic 
decay rate $1/\tau_{anharmonic}$ and the elastic scattering rate due to point defects $1/\tau_{defect}$. $1/\tau_{anharmonic}$ is given by Klemens formula 
$\gamma_{0}[1 + n(\omega_{TA}) + n(\omega_{LO})]$,\cite{Klemens66,Vallee91} where $\gamma_{0}$ is an effective anharmonic constant, $n(\omega)$ is the phonon distribution function, and $\omega_{TA}$ and $\omega_{LO}$ are the frequencies of the TA and the LO phonons, respectively. As pointed out earlier,\cite{Hase00} $1/\tau_{defect}$ is expected to increase linearly with increasing concentration of point defects, $1/\tau_{defect}$ = $\gamma_{\omega_{LO}}$$N_{d}$, where $\gamma_{\omega_{LO}}$ is a phonon-defect scattering constant and $N_{d}$ is the density of defects. 
In photo-excited GaP, the scattering by hot carriers (or photo-generated plasma) should also be taken into account; $1/\tau_{plasma}$ = $C_{LO}\omega_{P}^{2}$, 
where $\omega_{P}$ is the plasma frequency, $C_{LO} = (\omega_{LO}^{2} - \omega_{TO}^{2})/2\omega_{LO}^{4} \tau_{\infty}$ is a constant, 
where $\omega_{TO}$ is the TO phonon frequency, and $\tau_{\infty}$ is an average carrier momentum scattering time.\cite{Vallee97} 
Since $\omega_{P}^{2} \propto N$, where $N$ is the carrier density, $1/\tau_{plasma} \propto N$.\cite{Vallee97,Hase99} 
In total, the effective decay rate $1/\tau$ can thus be given by, 
\begin{eqnarray} 
\frac{1}{\tau} & = & \frac{1}{\tau_{anharmonic}} + \frac{1}{\tau_{defect}} + \frac{1}{\tau_{plasma}}.
\label{TAU}
\end{eqnarray}
From the literatures, we set $1/\tau_{anharmonic}$$\approx$ 0.08 ps$^{-1}$ (Ref. \cite{Ishioka15}), and thus $1/\tau_{plasma}$$\approx$ 0.04 ps$^{-1}$ is obtained from the total decay rate of $1/\tau$$\approx$ 0.12 ps$^{-1}$ before Ga ion irradiation. 
Since the longer dephasing time of 9.1 ps corresponds to $1/\tau$$\approx$ 0.11 ps$^{-1}$ [see the inset of Fig. \ref{Fig4}(a)], 
it is expected that the value of $1/\tau_{plasma}$ decreases by $\approx$ 0.01 ps$^{-1}$, 
which is obtained by the difference of $1/\tau$ before and after the ion irradiation, i.e., 0.12 ps$^{-1}$ - 0.11 ps$^{-1}$ = 0.01 ps$^{-1}$. In addition, because the anharmonic decay channel ($1/\tau_{anharmonic}$) depends mainly on the distribution of lower lying TA phonons and thus on the lattice temperature, the change of $1/\tau_{anharmonic}$ is negligibly small in the present study. 
We would argue that the term $1/\tau_{plasma}$ decreases over the $1/\tau_{defect}$ term in the case of GaP after the Ga$^{+}$ irradiation, 
i.e. $1/\tau_{defect}$ $<<$ 0.01ps$^{-1}$. 
This means that the concentration of carriers in the conduction band of GaP was reduced by the carrier trapping via defect states in the band-gap. 
\begin{figure}
\includegraphics[width=7.5cm]{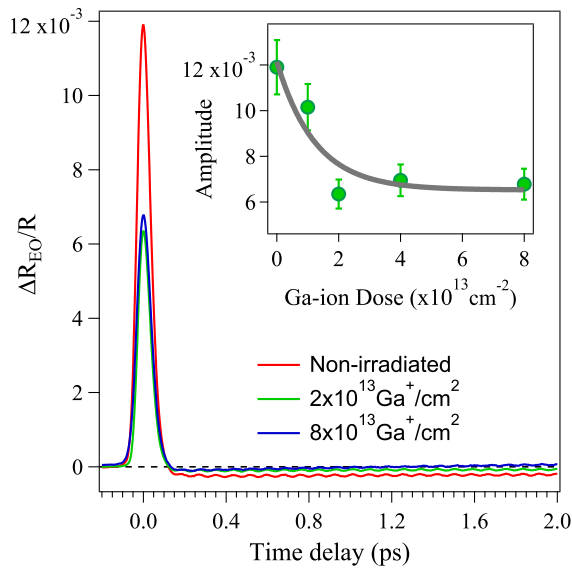}
\caption{The electronic-optic response from photoexcited GaP for non-irradiated, 2.0$\times$10$^{13}$Ga$^{+}$/cm$^{2}$, and 8.0$\times$10$^{13}$Ga$^{+}$/cm$^{2}$. The dashed line represents the zero level of the signal. 
The inset shows the amplitude of the electronic response at zero time delay as a function of Ga$^{+}$ dose. The solid curve is a guide for the eyes.}
\label{Fig5}
\end{figure}

\begin{figure}
\includegraphics[width=8.0cm]{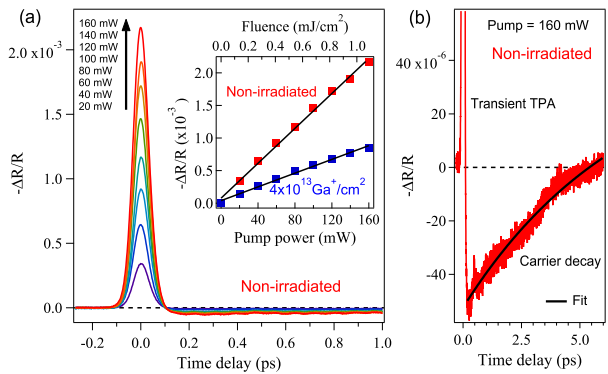}
\caption{
(a) The carrier response observed for non-irradiated GaP by isotropic reflectivity change ($\Delta R/R$) at various pump fluences. 
The dashed line represents the zero level of the signal. 
The inset shows the amplitude of the carrier response at zero time delay as a function of the pump fluence. The solid lines are the linear fit.
(b) The enlarged part of (a) for the pump power of 160 mW. The black solid line is the fit using a single exponential function with $\sim$ 3.5 ps decay time.
}
\label{Fig6}
\end{figure}

To test the above hypothesis of the dominant contribution for the prolonged dephasing time from $1/\tau_{plasma}$, we show the electro-optic response from GaP in Fig. \ref{Fig5}. 
The transient electro-optic response exhibits surprisingly ultrashort time scale, i.e., a full width at half maximum (FWHM) of $\approx$45 fs at the time delay zero, corresponding to nearly autocorrelation function of the laser pulse used ($\approx$25 fs), suggesting the transient signal is dominated by degenerate TPA from the one-pump and one-probe photons.\cite{Chong04} 
We found that the peak intensity of the transient electro-optic response in Fig. \ref{Fig5} decreases as the Ga$^{+}$ dose increases. This implies either a depletion of TPA by the deep levels associated with Ga$^{+}$ ions or ultrafast trapping of photogenerated carriers by the deep levels. 
In photoexcited GaP, the photogenerated electrons decays via intervalley scattering from the zone center ($\Gamma$ point) into the lower-lying $X_{1}$, $X_{2}$ and L valleys within only $\approx$30 fs.\cite{Sjakste07} Therefore, the ultrafast electro-optic response ($\approx$45 fs) can be explained by the intervalley scattering just after the TPA, 
rather than the trapping by deep levels, which usually takes more than several hundreds femtoseconds. 
The carrier density excited via TPA with the one-pump and one-probe photons can be expressed by,\cite{Sabbah02,Hutchings92}
\begin{equation}
N_{TPA} = \frac{\beta I_{pump}I_{probe}}{2E_{p}\tau_{p}}, 
\label{TPA}
\end{equation}
where $\beta$ = 2.0 cm/GW (for 830 nm) represents the TPA coefficient,\cite{Grinblat19} $I_{pump}$ $\approx$ 1.19 m$\mathrm{J/cm^2}$ ($I_{probe}$ $\approx$ 60 $\mathrm{\mu J/cm^2}$) is the laser fluence of the pump (probe) pulse, $E_{p}$ = 1.49 eV is the photon energy of the laser used, and $\tau_{p}$ = 25 fs is the pulse length.
The carrier response measured for various pump fluences by isotropic reflectivity change ($\Delta R/R$ $\propto$ $N_{TPA}$)\cite{Motojima19} revealed the linear dependence on $I_{pump}$ as shown in Fig. \ref{Fig6}(a), indicating the relationship given by Eq. (\ref{TPA}). 
By using the linear fit to the fluence dependence in the inset, we found the $\beta$ decreased by 40\% after irradiation. So we may estimate the $\beta$ after irradiation to be 0.8 cm/GW; 
this is possible since $\beta$ depends on refractive index \cite{Hutchings92} and it would be changed after irradiation as revealed by changing the color of the irradiated area [Fig. \ref{Fig1}(b) inset]. 
We thus estimated the maximum photo-generated carrier density via TPA to be $N_{TPA}$ $\approx$1.2$\times$10$^{16}$/cm$^{3}$ before ion irradiation. 
We note that the slow carrier decay signal, appearing just after the transient TPA in Fig. \ref{Fig6}(b), would represent signal from the carrier plasma scattered from the $\Gamma$ point to the $X_{1}$, $X_{2}$ and L valleys, the magnitude of which was decreased after ion irradiation (data not shown but also visible in Fig. \ref{Fig5}). 
Note also that the carrier plasma would interact with coherent LO phonons via Fr\"{o}lich coupling,\cite{Ishioka15} 
i.e., the effect of excited carriers through one-pump and one-probe photons induced TPA on the dephasing of the coherent LO phonon was examined by probe-power dependence on the phonon dephasing, which exhibited the decrease of the time constant with increasing the probe fluence (data not shown). 
The decrease in the TPA excited carriers will therefore weaken the plasmon-LO phonon coupling after the intervalley scattering, i.e., after $\sim$ 30 fs, resulting in the increase of the dephasing time of the coherent LO phonon (Fig. \ref{Fig4}). 

\begin{figure}
\includegraphics[width=7.0cm]{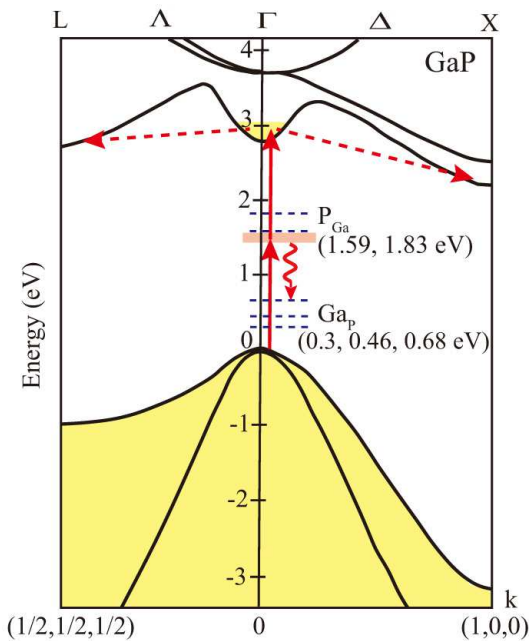}
\caption{The schematic electronic band structure of GaP, based on the energy scales in Ref. \cite{Zallen64}. 
The solid arrows show the two-photon absorption from the valence band and the dashed arrows represent subsequent scattering into the X- and L-valleys, while the wavy arrow indicates a possible decay or trapping of carriers via P$_{Ga} \rightarrow$ Ga$_{P}$ bands. 
}
\label{Fig7}
\end{figure}

In the present case, the energy of the pump photon ($E_{p}$=1.49 eV) approaches the defect-induced deep levels as an intermediate-state. In fact, the anti-site defect of P$_{Ga}$, where one Ga site is occupied by a P atom, shows the mid-gap energy level of 1.59 eV from the top of the valence band,\cite{Puska89} 
 which will not allow the pump photon to exactly resonant at the P$_{Ga}$ level, but partly contribute to one-photon absorption at the P$_{Ga}$ bands (see Fig. \ref{Fig7}).\cite{Zallen64} 
Thus, we can conclude that the excited carrier plasma density at the zone center ($\Gamma$ point) can be decreased by depletion of TPA by nearly resonant one-photon absorption at P$_{Ga}$ states, whose carriers decay into underlying Ga$_{P}$ band. 

In summary, we observed the prolonged dephasing time of the coherent LO phonon in Ga ion irradiated GaP by use of the fs pump-probe 
reflectivity technique. The prolongation of the dephasing time of the coherent LO phonon observed for defective GaP is not intuitive, 
since defects generally promote dephasing. However, we present the effective model for suppression of interaction between photo-generated carrier plasma and coherent LO phonon  caused by quenching of the TPA photo-excited carriers by P$_{Ga}$ deep levels. Our observation 
of prolonged dephasing time, which may be called {\it phonon cooling} by lattice defects, provides a new insight into controlling the electrical properties of semiconductors for the development of device applications. In addition, the results obtained may be useful for investigating a possible generation of micro- or nano-structures of GaP and even other nonlinear materials by FIB technique, while the lattice properties unchanged. 

\section*{Acknowledgement}
This work was supported by CREST, JST (Grant Number. JPMJCR1875), Japan. T. I. acknowledges the support from Innovation School, General Affairs Headquarters, AIST, Japan. 
We acknowledge Prof. Takashi Sekiguchi of University of Tsukuba for helping the ion implantation using FIB. 

\section*{DATA AVAILABILITY}
The data that support the findings of this study are available within this article.
%\newpage

\pagebreak
%\newpage
%FIGURE CAPTIONS

\end{document}